# On coexistence of decentralized system (blockchain) and central management in Internet-of-Things

Hiroshi Watanabe, *National Yang Ming Chiao Tung University, hwhpnabe@gmail.com*

*Abstract*—Network is composed of logical nodes and edges for communications. Atomistic component of things connected to the network is a memory chip. Accordingly, the unique linkage of a memory chip and a logical node can be a promising to resolve the root-of-trust problem on the Internet-of-Things. For this aim, we propose a protocol of challenge-response using a memory chip.

For the central management, a central node controls the entry of electronic appliances with a memory chip into the network, and excludes a fake node (e.g., the spoofing entity) from the network that the central node manages. For the decentralized communications, Merkle's tree turns out being composed of memory chips to which the logical nodes are uniquely linked, respectively. The root of Merkle turns out being the memory chip that stores the latest record of data transaction. We can register this memory chip as a new block by satisfying the requirement of the proof-of-consensus. After blocks are chained, it gets harder for even the central node to manipulate transaction record among memory chips. By this way, the decentralized system (e.g., blockchain) and the central management can coexist. A new idea of security state is also discussed briefly.

*Index Terms*—Blockchain, Merkle's tree, central management, decentralized system, memory chip, IC chip.

## I. INTRODUCTION

Decentralized systems (e.g., blockchain) can maximize the value of the network applications. By uploading hardware to the network, the Internet-of-Things (IoT) can heighten the value of the network applications. However, the entry of hardware to the network should be under the central management for the authentication reason. The decentralized system and the central management should therefore coexist in the IoT network, though it has not been discussed enough.

What is the upload of hardware to the network? --- The network comprises logical nodes and communication edges (lines) connecting logical nodes. In general terms, the upload of hardware to the network is to uniquely link a physical entity to a logical node. The uploaded physical entity is a physical node. It is an electronic appliance whose atomistic component is an integrated circuit (IC) chip, in particular, a memory chip in the Neumann-type computers. The falsification of the linkage of a physical node and a logical node is the spoofing. The proof of no spoofing is the root-of-trust. The upload of hardware (i.e., the linkage of a physical node and a logical node) without the root-of-trust causes a kind of oracle problem. Because one may be forced to trust it with no proof and any existing Blockchain cannot resolve it. A practical reinforcement of Blockchain is therefore necessary to use Blockchain on the IoT network.

A logical node is allocated with an address on the network. This address is public on the network, such that an arbitral entry in the network can know it to reach the logical node. Following the concept of Diffie-Hellman [1], we can think that a public key can play a role of (public) address on the network. Allice has distributed her public key on the network. Anyone who entries to the network can receive her public key. Both Bob and Mike can encrypt their messages by using Allice's public key. They can distribute (or expose) their messages to her on the network. Any entry can receive those messages but only Allice can read the messages by decrypting them using her secret key that is not distributed on the network. While only Allice can read the messages, her public key can play a role of Allice's address on the network, which is public on the network. Like this, the atomistic component of a physical node linked to a logical address is a memory chip and a public key can play a role of address of the logical node (named, logical address). The root-of-trust is therefore the proof of the unique linkage of a memory chip and a public key.

Data can be transferred from a logical node to another, which data transfer can be denoted by an arrow. Suppose that a logical node can receive data transferred from plural logical nodes. Plural data transfers result in a tree diagram, whose root is the destination having all transfer records as well as the latest one. This is called Merkle's tree and its root is called the root of Merkle [2]. If we can prove the unique linkage of a memory chip and a logical node (represented by its public key) in some way, then it turns out being Merkle's tree of memory chips. The Merkle root is thus a memory chip which stores the latest transaction record of data transfer. In the network, miners look for a root of Merkle (represented by its public key) and then register it as a new block by satisfying the requirement for the proof-of-consensus. Accordingly, the registered new block is a memory chip having the latest transaction record and uniquely linked to the public key of the registered root of Merkle. By repeating this procedure, plural blocks are serially registered to construct the blockchain of memory chips with the root-of-trust (i.e., the unique linkage of memory chips and public keys).

A central node can control the upload of an electronic appliance (hardware) with a memory chip to the network. A logical node uniquely linked to this memory chip is under the central management by the central node. If the central node permits the upload of this electronic appliance, then the logical node uniquely linked to the memory chip mounted in this electronic appliance is permitted to entry to the network. If the central node denies the upload of this electronic appliance, then the logical node uniquely linked to the memory chip mounted in this electronic appliance is denied to entry to the network. But, after the blocks of memory chips are chained, the central node can hardly manipulate transaction record in the blockchain of memory chips that the central node has permitted to entry. Because it must be necessary to redo the proof-of-consensus for the chained blocks [3]. As the blockchain length increases, it becomes harder. That is, the central node can control the entry of an electronic appliance (represented by its memory chip) to the network but can hardly manipulate the transaction record among memory chips that the central node has permitted to entry. By this way, the decentralized system and the central management can coexist with the root-of-trust (i.e., the unique linkage of memory chips and public keys).

In this work, we illustrate a conceptual solution for this in as a clear manner as possible. In II, we describe the method to realize the unique linkage of memory chips and public keys. In III, we illustrate the blockchain of memory chips. IV and V are devoted to discussion and summary, respectively.

## II. Unique linkage of memory chips and public keys

This can be realized by the challenge-response protocol (CRP) using a memory chip and a physical random number (PRN) which is specific to the memory chip [4]. Fig. 1 illustrates the challenge-response (a) without and (b) with the spoofing. The connected device A inspects the connected device B by asking "Hey B, who are you?" (Challenge C). In (a), without the spoofing, the connected device B replies to the connected device A, "I am logical address B" (Response R). However, without the root-of-trust, we are unsure if the connected device B and the logical address B are really linked. In (b), a hacker spoofs the logical address B, to "Hey B, who are you?" (Challenge C), the response is "I am logical address B" (Response R). That is, the CRP without the root-of-trust makes nonsense in the IoT network.

### A. Root-of-Trust using a memory chip

First, suppose that a public key can play a role of a logical address B. Next, suppose that a secret key is irreversibly generated from a code specific to a memory chip which is an atomistic component of a physical node (e.g., connected device B). Using some algorithm for the public key infrastructure (PKI), we can uniquely link the secret and public keys. Using this specific code and the PKI, we can uniquely link the memory chip and the logical address B.

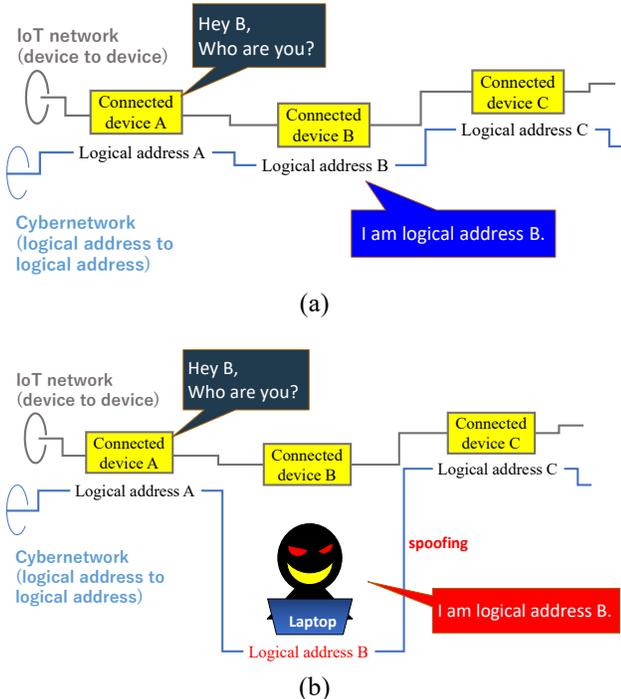

Fig. 1. Challenge-Response Protocol. (a) without spoofing. (b) with spoofing.

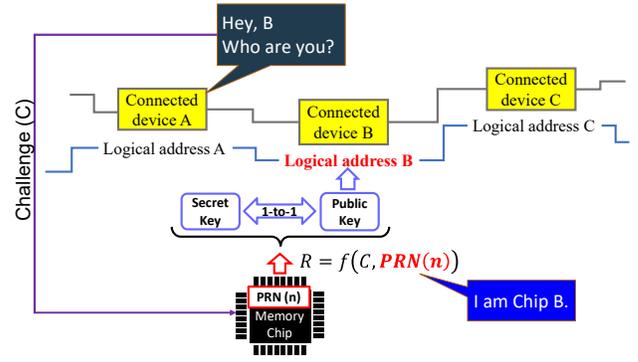

Fig. 2. CRP with physical randomness specific to a memory chip.

In general, we can obtain the response ($R_n(C)$) from the physical random number of chip ($n$), $PRN(n)$, and the challenge (C) using a function, $f$, as follows.

$$R_n(C) = f(C, PRN(n)) \qquad (1)$$

In Fig. 2, the connected device A sends the challenge (C) "Hey B, who are you?" to the memory chip ($n$) having $PRN(n)$ of the connected device B. The response (R) generated using (1) is "I am chip B." If the connected device B was spoofed by hacker's laptop, the response would be "I am a chip in hacker's laptop". The CRP using a memory chip makes sense for the cyber-attacks. Because no cyber-attack can tear off memory chip from the motherboard of the connected device B.

### B. PRN and PKI

We can generate one or two prime number(s) from this $R_n(C)$ in some way. Thus, we can generate a "uniquely linked" pair of secret and public keys using an algorithm of PKI --- RSA [5], ElGamal [6], etc. Therefore, the secret key (SK) and public key (PK) of chip ($n$) are respectively written as follows.

$$SK_n(C) = g(R_n(C)) \qquad (2)$$
$$PK_n(C) = \bar{g}(R_n(C)) \qquad (3)$$

The ($g, \bar{g}$) are the key generation functions. It is preferable that prime numbers are great enough. The (3) connects chip ($n$) and public key $PK_n(C)$ through (1). Below we evaluate randomness to validate if $PRN(n)$ is specific to chip ($n$) in practice.

### C. A concrete example of retrieving PRN from memory chips

Fig. 3 illustrates the redundancy mechanism of a memory chip [7]. There are several blocks on chip. Each block is composed of many integrated memory cells (corresponding to bits), which are arrayed in the X (Row)-Y (Column) plane. It is further divided into two cell arrays (redundancy and regular arrays). The Y-decoders A and B control the row access in the redundancy and regular arrays, respectively. An access code is input to a peripheral circuit to choose either Y-decoder A or B or both. Since memory chip is a mass-product, it is impossible to exclude all failures from the cell arrays perfectly. Suppose there are plural failure bits in the regular array, wherein the row number with a failure bit is denoted by $F_i$ for $i = 1, 2, \ldots m$ with $m$ being the number of rows with a failure bit in the regular array. While this $m$ is small enough, that is, the failure is controllable under the manufacturing specification, we can apparently exclude the failure bits from the memory access operations (i.e., read, write, and erase in the regular array). Inputting an access code into the peripheral circuit, we can choose a normal access mode to use only Y-decoder B. We, thus, access the addresses from the top of the regular array along a column ($j$) chosen by X-decoder. When arriving at $F_i$, we swap the bit with another bit on the row number $R_i$ in the redundancy array. If the bit at the address $(R_i, j)$ is not a failure bit, we can apparently exclude the failure bit at $(F_i, j)$ from the memory access operations in the regular array. In usual, the users of DRAM can access only the regular array. As $m$ decreases, we can reduce the necessary number of rows in the redundancy array. It can suppress the probability that a failure bit is found in the redundancy array as well. As improving the manufacturing quality, the average of $m$ gets smaller. We can expect that $m$ is small enough in the mass-produced memory chips, but it can hardly be zero. In general, the distribution of rows with a failure bit in the regular array of chip ($n$), denoted by $\{F_i\}_n$, is out of the manufacturing control. It can cause a physical randomness specific to chip ($n$). The inner memory stores this $\{F_i\}_n$ to exclude failure bits from the memory access operation. Below we illustrate the method to retrieve $PRN$ ($n$).

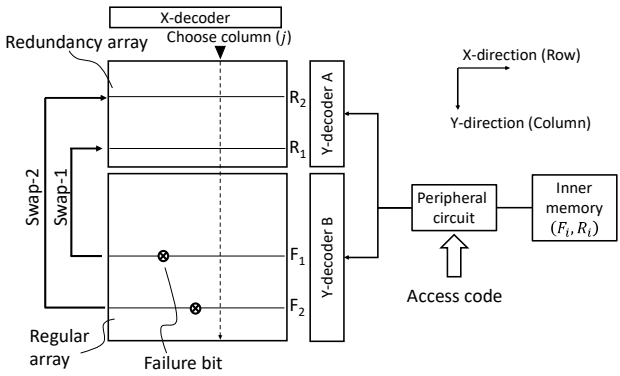

Fig. 3. Redundancy mechanism.

*Preprocess)* We input a normal access code to the peripheral circuit to use Y-decoder B. This is the normal access mode. We write state-0 along the $j$-th column (chosen in advance) in the regular array. See Fig. 4 (a). Next, inputting a special access code to the peripheral circuit, we change the access mode to the special access mode to use Y-decoder A. Then, we write state-1 along the $j$-th column in the redundancy array. See Fig. 4 (a).

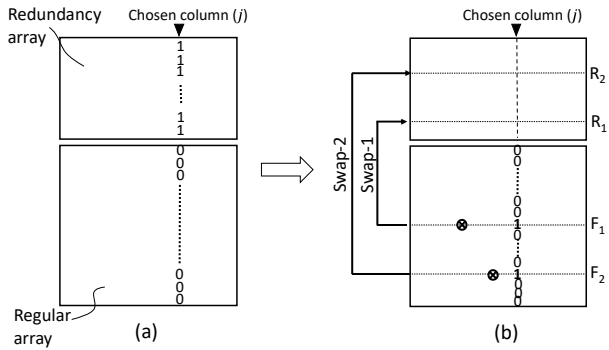

Fig. 4. Retrieve PRN from memory chip. (a) preprocess. (b) read.

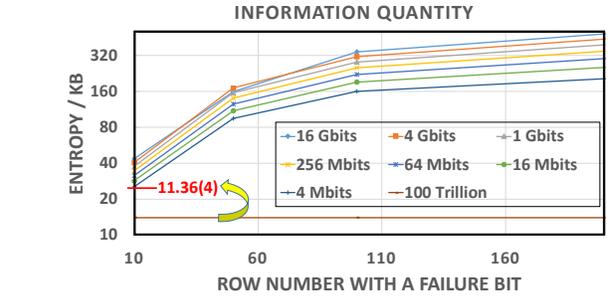

Fig. 5. Entropy of randomness with memory chips

*Read)* In a normal access mode (using Y-decoder B), we read bits along the $j$-th column in the regular array. By the swapping, readout turns out being state-1 on rows with a failure bit ($F_i$) and state-0 on the other rows. See Fig. 4 (b). Thus $\{F_i\}_n$ specific to chip ($n$) turns out being a physical random number, $PRN(n)$.

*PRN Stability)* Reliability of the inner memory storing $\{F_i\}_n$ determines that of $PRN(n)$. In general, the inner memory (e.g., a fuse memory) is more stable to the environmental change than cell arrays. In experiment using the dynamical random access memory (DRAM) chips, that is the most popular IC product for the main memory, the excellent characteristics of retention and stability to temperature change were clearly shown [4], [7].

*Randomness)* As an example, we choose 4 Mbits - 16 Gbits (i.e., the old and new generations) DRAM for evaluating the entropy of chip-specific randomness. For the ease of discussion, we assume that $m$ is much smaller than the total row number ($Y$), and the row and column numbers are the same (i.e., isotropic layout) in each block. The number of blocks on a chip is thus $L^2$. For $L = 1$, the entire cell array on chip is one block. The $Y$ is in the order of 100,000 for 16 Gbits, which is most advanced recently. The randomness can be estimated by calculating the combination product of $Y$ and $m$, i.e., $_YC_m$. For an arbitral $L$, the randomness can be estimated by $_{L\times Y}C_m$. Following Boltzmann, the entropy divided by $k_B$ is $\ln {_{L\times Y}C_m}$, where $k_B$ is the Boltzmann constant. For $L = 1$, although it is an underestimation a little bit, we plot information entropies of PRN from DRAM chips (4 Mbits to 16 Gbits) in Fig. 5. For $m = 10$, $\ln {_{L\times Y}C_m} > 10^{25}$ even for 4 Mbits with $L = 1$ (the worst case in this estimation). If one hundred trillion ($10^{14}$) IoT devices will be deployed all over the world, the probability that two chips have a same PRN is less than $10^{-11}$ (i.e., 0.00001 ppm). Thus, the randomness is good enough in the 4 Mbits – 16 Gbits products of DRAM. In addition, a couple of ten rows ($\gg m$) in the redundancy array is good enough for swapping to hide most of failure bits from the memory access operations. The redundancy array comprising twenty rows is much narrower than the regular array ($20 \ll 2,000$ and $100,000$ in 4 Mbits and 16 Gbits products, respectively). It does not matter in the rough estimation of the entropy even though increasing $L$. Therefore, we can obtain a sufficient randomness to retrieve $PRN$ with no additional area for cell array. It tells us there is no chip area penalty. The present method is feasible to the wide generations of the existing DRAM products (4 Mbits – 16 Gbits). It is self-evident that the present method is also feasible to the coming generations beyond 16 Gbits.

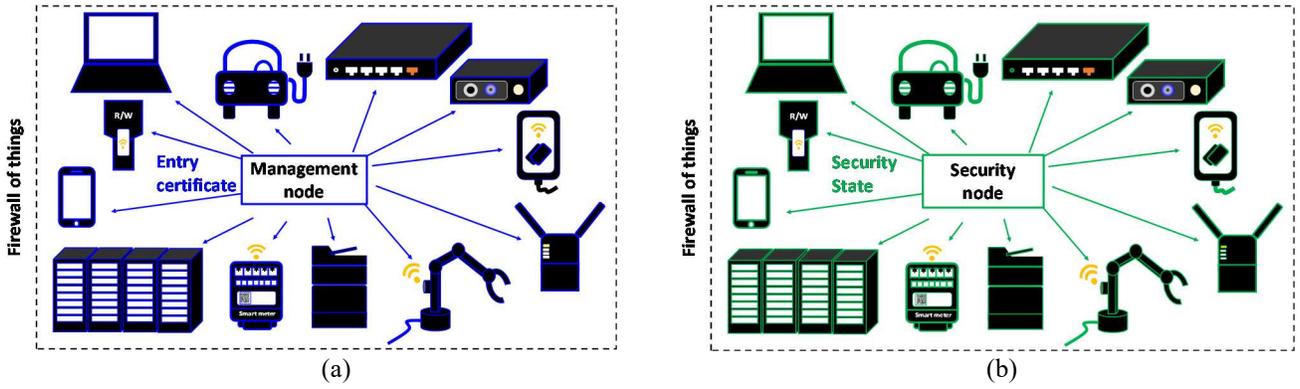

Fig. 6. Central management. (a) by management node. (b) by security node.

*D. Independencies of chips*

While the entropy of randomness specific to a memory chip is good enough, we can hardly retrieve a same random number which is specific to two different chips, as discussed above.

$$PRN(n1) \neq PRN(n2) \text{ while } n1 \neq n2 \quad (4)$$

While satisfying (4), we respectively convert (1) – (3) to:

$$R_{n1}(C) \neq R_{n2}(C) \text{ while } n1 \neq n2 \quad (5)$$
$$SK_{n1}(C) \neq SK_{n2}(C) \text{ while } n1 \neq n2 \quad (6)$$
$$PK_{n1}(C) \neq PK_{n2}(C) \text{ while } n1 \neq n2 \quad (7)$$

Furthermore, we can write down as follows.

$$R_n(C_1) \neq R_n(C_2) \text{ while } C_1 \neq C_2 \quad (8)$$
$$SK_n(C_1) \neq SK_n(C_2) \text{ while } C_1 \neq C_2 \quad (9)$$
$$PK_n(C_1) \neq PK_n(C_2) \text{ while } C_1 \neq C_2 \quad (10)$$

The (5) – (10) characterize functions that are necessary for memory chips to satisfy the root-of-trust (i.e., the unique linkage of memory chips and public keys). In Fig. 6 (a), physical nodes satisfying these configures firewall of things [8].

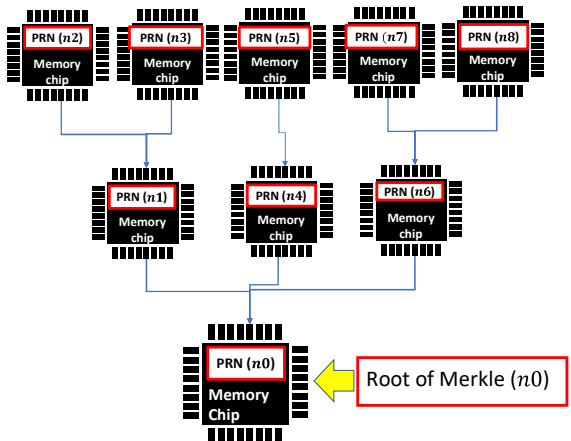

Fig. 7. Merkle tree of memory chips.

*E. Auto-detection and auto-remove (Entry control)*

Auto-detection and auto-remove of the fake and vulnerable nodes can be performed using (5) – (10) and the smart contract. Details will be discussed elsewhere.

*F. Security State*

For each $n$, we generate response, secret and public keys using a challenge $C_l$, where $l$ is a natural number. Regard the generation of them as appearing similar to an observation in quantum mechanics (denoted by $\hat{R}$, $\hat{S}$ and $\hat{P}$, respectively). To deduce (8) – (10), we can begin with the following equations, respectively.

$$\hat{R}_n|l\rangle = R_n(C_l)|l\rangle \quad (11)$$
$$\hat{S}_n|l\rangle = SK_n(C_l)|l\rangle \quad (12)$$
$$\hat{P}_n|l\rangle = PK_n(C_l)|l\rangle \quad (13)$$

The $R_n(C_l)$, $SK_n(C_l)$ and $PK_n(C_l)$ of chip $(n)$ can describe a security state denoted by $|l\rangle$ (like an eigenstate in quantum mechanics). However, the security state of a physical node $(n)$ can be determined by choosing a challenge $C_l$. Since the state $|l\rangle$ has been chosen before the observation (i.e., the generation of response and secret and public keys), (11) – (13) are closer to Einstein's hidden variables [9] than Schrödinger's cat. Anyway, this can confuse adversaries. Fig. 6 (b) illustrates the merit of security state. In the network, another central node (security node) can change response and secret and public keys of any physical nodes by replacing $C_l$ periodically or at his convenience. It is effective when some kind of security problems is found. Details will be discussed elsewhere.

The $C_l$ used by a security node should differ from challenge used by a management node. The entry and the security state are controlled by two independent central nodes (management and security nodes, respectively). This reminds us of the separation of three powers in the democracy, even though this is a central control indeed.

III. BLOCKCHAIN OF MEMORY CHIPS

Most of physical nodes is a Neumann computer which is composed of an input-output (I/O), an arithmetic unit and a memory chip. Data, stored in a memory chip of a physical node, is retrieved and processed by the arithmetic unit and then output from I/O to the external. This data is received by I/O of another physical node and then processed by the arithmetic unit and then stored in the memory chip of the receiving physical node. That is, data is really transferred between memory chips.

*A. Merkle tree of memory chips*

In Fig. 7, we suppose that data stored in chip $(n0)$ has been transferred from chips $(n1, n4,$ and $n6)$ with recording the updated history of data transfer to chip $(n0)$. Each data transfer is modeled in Fig. 8, which illustrates that from node $(n-1)$ to node $(n)$. In chip $(n)$, the $R_n(C_l)$ having been obtained from $C_l$ and $PRN(n)$ at security state $|l\rangle$ according to (1) is used to generate $SK_n(C_l)$ and $PK_n(C_l)$ using (2) and (3), respectively. While (4) is satisfied, we regard (5) – (13) as all satisfied. The chip-specific $R_n(C_l)$ is related to $SK_n(C_l)$. The unique linkage of memory chip and public key is thus assured by the unique linkage of secret and public keys that are generated using a same chip-specific $R_n(C_l)$. We can regard logical node $(n)$ as equal to physical node $(n)$ on the network. We can combine them to call it as node $(n)$, if not specially noted.

In the node $(n-1)$, the $PK_{n-1}(C_l)$, hash value $(n-2)$, and electronic signature $(n-2)$ are converted to a new hash value $(n-1)$ using a hash function (e.g., SHA256). Subsequently, the node $(n-1)$ gets $PK_n(C_l)$ on the network. (Public key is public on the network.) The got public key and the generated hash value $(n-1)$ are encrypted to be a new electronic signature $(n-1)$ using $SK_{n-1}(C_l)$. The node $(n-1)$ sends the hash value $(n-1)$ and the electronic signature $(n-1)$ to the node $(n)$. This transfer is as same as used in the existing blockchain if we exclude memory chips from the illustration (drawn in red). That is, our solution for the unique linkage of memory chips and public keys is fully compatible to the existing blockchain.

Move back to Fig. 7. Data stored in chip $(n1)$ has been transferred from chips $(n2$ and $n3)$ with recording the updated transferring history in chip $(n1)$. Data stored in chip $(n4)$ has been transferred from chip $(n5)$ with recording the updated transferring history therein. Data stored in chip $(n6)$ has been transferred from chips $(n7$ and $n8)$ with recording the updated transferring history therein. It turns out being a tree of Merkle. There is chip $(n0)$ at the bottom, named, the Merkle root $(n0)$,

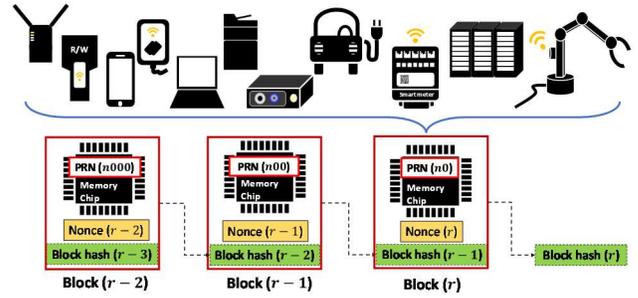

Fig. 9. Blockchain of memory chips.

with recording the entire history of data transfers from plural chips to the root chip $(n0)$.

*B. Blockchain of memory chips*

In Fig. 9, we follow the conventional mining process to chain blocks. A miner can register the newest block hash $(r)$ by hashing an added nonce $(r)$, a stamp of the root chip $(n0)$ (e.g., $PK_{n0}(C_l)$, hash value $(n0-1)$, etc.), and the previous block hash $(r-1)$, such that the newly generated block hash $(r)$ satisfies the requirement for the poof-of-consensus by tuning nonce $(r)$. The block hash $(r-1)$ was registered by hashing an added nonce $(r-1)$, a stamp of the root node $(n00)$ (e.g., $PK_{n00}(C_l)$, hash value $(n00-1)$, etc.), and the previous block hash $(r-2)$, so that the block hash $(r-1)$ satisfies the requirement for the poof-of-consensus by tuning nonce $(r-1)$. Repeating this procedure, the miners have constructed the blockchain of memory chips.

There are plenty of diverse types of electronic appliances (physical nodes) --- routers, PCs, smartphones, card readers, RFID reader-writers (R/Ws), surveillance cameras, industrial robots, smart meters, printers etc. in Fig. 9. However, DRAM is a general-purpose memory and installed in most of electronic appliances. In Fig. 10, any nodes are approved to communicate each other without being monitored by any central nodes inside the firewall of things [8]. One of central nodes (management node) controls the entry of physical nodes to the network and prohibits the spoofing by a hacker's laptop with non-registered chip (see Fig. 1 (b) as well). Another central node (security node) controls security state. The unique linkage of memory chips and public keys is fully compatible to the blockchain construction.

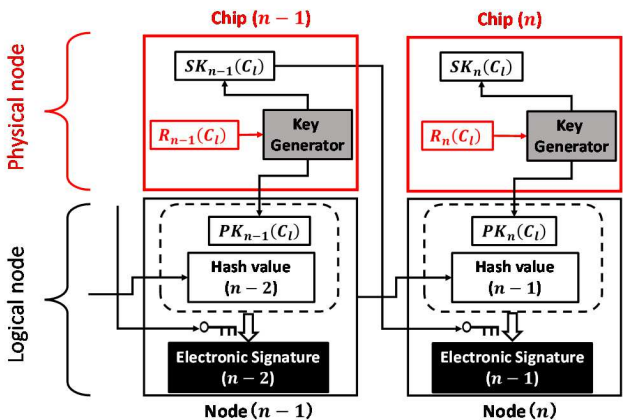

Fig. 8. Data transaction. See it by removing red ones.

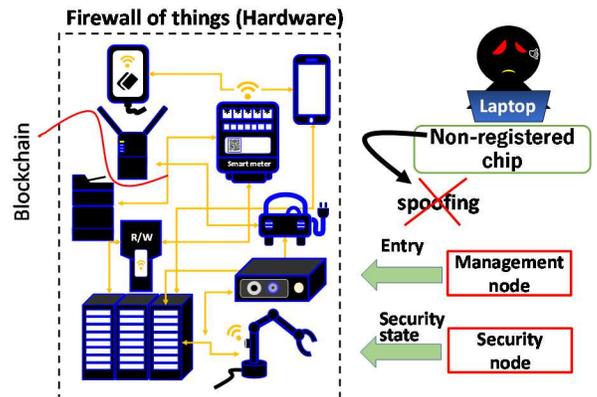

Fig. 10. The coexistence of central management and decentralized system (blockchain).

Blockchain protects communications in the firewall of things. This is the decentralized communications under central control.

*C. Resilient Blockchain of things*

There is no risk of mechanical failures in logical accounts. But, in the IoT network, there is the risk of mechanical troubles. If we replace a defected device having a memory chip ($na$) by a new device having a memory chip ($nb$) in the blockchain of memory chips, the response and public and secret keys will change since $PRN(na)$ and $PRN(nb)$ differ. It requires the reproduction of the Merkle root using the entire tree. Another case to require the reproduction of the Merkle root is the change of security state for some security reason.

However, we can follow the idea of Beyond Blockchain one (BBc-1) [10], which can be regarded as resilient blockchain. Only limited subtrees are good enough to reproduce the Merkle root. It can reduce the reproduction cost. The reproduced tree can be registered as a new block after the proof-of-consensus.

## IV. Discussion

Since the version of security of IoT devices vary greatly from old to new, a hacker may get an authenticate information of a vulnerable IoT device. Any advanced security parts (PUC, protected memory [11] etc.) cannot protect old IoT devices having been shipped with the security parts in the former generations or with no security parts.

The present method can be an application software, which retrieves PRN from existing IC chips in the distributed IoT devices. It was demonstrated that (4) can be satisfied within the quality control of mass-produced DRAM chips (4 Mbits – 16 Gbits), with no chip area penalty. Most of existing IoT devices (old and new) mounts a DRAM chip with the bit capacity being 4 Mbits – 16 Gbits. To adopt the present method, we may install the application software and not replace old IoT devices by new ones with the advanced security parts (PUC etc.).

Smartphone suppliers can construct the unique linkage of smartphones and public keys using the present method. This is the firewall of smartphones. End users are required to download an application to retrieve PRN from DRAM chip existing in their smartphones but not to replace their old smartphones with new ones so that their smartphones can enter into the firewall.

In the industrial manufacturing, the supervisory control and data acquisition (SCADA) is collecting the attention. In the semiconductor manufacturing, for example, a medium class factory with ~40,000 wafers per month has about 3,000 units of equipment. If each unit has 100 parameters in average, then about 300,000 parameters are to be optimized even in a medium class factory. Twice or more in a huge class factory. In the SCADA, those parameters are collected from plural smart factories by IoT sensors. A remote facility of AI dynamically and consistently optimizes data (parameters) collected from the plural smart factories. Therefore, there is an increasing risk that the cyber-attack will upset the global supply chain and theft technical know-how. Since each IoT sensor has a DRAM chip, we can construct the unique linkage of IoT sensors and public keys over the remote-controlled smart factories distributed globally. This is the firewall of IoT sensors and industrial robots.

In the logistics control, the security of RFID tags has been discussed. However, a spoofed R/W can deliver parcels to an illegal destination. Thus, we propose the unique linkage of R/Ws and public keys [12]. This is the firewall of R/Ws.

For the traceability in the supply chain of IC chips, we can construct the firewall of IC chips using the present method.

## V. Summary

Taking into account the quality control of mass-produced memory chips, i.e., atomistic component of IoT devices, we can construct the unique linkage of memory chips and public keys with a full compatibility to the existing blockchains. The central management of IoT devices and decentralized communications among IoT devices can therefore coexist. The present method is valid to various IC products as well as DRAM. Application range is pretty-wide and not limited to discussed here.

Acknowledgment

The author thanks to T. Hamamoto, Y. Nagai, J. Liang, M. Chang, E. Tseng, J. Moon, K. Saito, K. Taniguchi, S. Torisawa, T. Kato, KY Tsai, J. Chen, L. Chang, Y. Hirota, A. Kinoshita for valuable and stimulating discussions, and T. Okada, S. Miyazaki, and H. Fukuyama for support on experiment.